\documentclass[onecolumn,amsmath, amssymb, nobibnotes, aps, prf, floatfix, showpacs]{revtex4-2}

\usepackage{graphicx,wasysym,color,booktabs}

\usepackage{epstopdf}
\usepackage{graphicx}
\usepackage{xcolor}
\usepackage[latin1]{}
\usepackage{tabularx}
\usepackage{placeins}
\usepackage[separate-uncertainty=true,multi-part-units=single]{siunitx}
\usepackage{hyperref}

\usepackage{soul}

\begin{document}

\title{Novel microfluidic strategy for the production of sodium alginate fibers with regular inclusions at very high throughput}

\author{Francesco Marangon}
\author{David Baumgartner}
\author{Carole Planchette}
\email[]{carole.planchette@tugraz.at}
\affiliation{Institute of Fluid Mechanics and Heat Transfer, Graz University of Technology, A-8010 Graz, Austria}
\date{\today}

\begin{abstract}
 Scalable technologies for the production of bio-compatible complex microfibers with controllable size and composition at competitively high throughput are urgently needed in order to meet the growing demand for such microstructures in pharmaceutical or biomedical applications. Here, we introduce a new in-air microfluidic strategy with throughput of $> 1400 ml/h$ (corresponding to $> 17000\, m$ fiber per $h$). The microfibers of uniform diameter have regular inclusions, which can potentially be used for encapsulating cells into a protecting and nutrient environment, or for finely tuning the release of various actives at individualized doses. With the help of a newly developed prototype, we test seven different liquid combinations and obtain seven types of fibers, whose "dry" diameter range between $73\mu m$ and $141\mu m$. The principle of our approach is to solidify the complex liquid structures generated by the controlled collisions of a drop stream with a continuous liquid jet, in air, via ionic crosslinking. After the stream of water-based droplets, which constitute the inclusions, collides in-air with the alginate-based jet (jet 1), the generated \textit{drops-in-jet} compound is brought in contact with a second jet (jet 2) containing divalent cations ($Sr^{2+}$ or $Ca^{2+}$) to initiate the solidification. Finally, the fibers are collected via a horizontal spinning plate, let to dry, i.e. to fully equilibrate under controlled conditions, and characterized by their elongation at break and Young's modulus. 

\end{abstract}

\begin{description}
\item[PACS numbers]
\end{description}

\maketitle

\section{Introduction}
\label{sec:introduction}

More than two decades ago, as the field of microfluidics emerged, the control and manipulation of very small amounts of fluid volumes (nano-, micro-, millimeter scale) became possible and has since then gained more and more relevance \cite{ref:Lei2018, ref:Castillo2015}. To date and thanks to the small amounts of fluids at stake, most of the technical applications of microfluidics are related to biomedical, pharmaceutical or chemical industries \cite{ref:Ohno2008, ref:Cui2019}. One aim of microfluidic systems is to propose a solution, which takes over the sampling, processing and result displaying. Tasks, which are usually executed combining several laboratory instruments, should be performed within one single small device. Microfluidic devices are, therefore, often called biochip, lab-on-chip or micro total analysis system \cite{ref:Hardt2007, ref:Castillo2015}. In parallel to this development, another branch of microfluidics, often called droplet-based microfluidics, has been focusing on the generation and manipulation of microdroplets \cite{ref:Seemann2011}. One motivation is to take advantage of the high ratios of surface area to volume offered by microdroplets, a characteristic which is of interest in many chemical applications. Another aspect is the fine tunability and precise handling associated to microdroplets. The latter can thus be used in several encapsulation processes to obtain spherical capsule precursors, possibly showing a core-shell structure with for example, encapsulated cells \cite{ref:Clausell2008, ref:Gong2009, ref:Feng2019}. Despite the advantages and the great potentials of droplet-based microfluidics, there are still several challenges to tackle before a successful and broad usage in industries can be established \cite{ref:Kaminski2016, ref:Chiu2017, ref:Zhou2020}. 

The wetting and cross-contamination problems observed with the encapsulation of microorganisms constitute two of these critical points \cite{ref:Hao2019, ref:Kaminski2016}. In most cases, the cells are embedded in droplets made of aqueous solutions, while the carrying phase usually consists of an immiscible liquid, like silicone oil. Whereas, at first glance, the passage of cells from one droplet to another one seems impossible, such events occur, which can have dramatic consequences. Cross-contamination indeed happen via different processes: (i) coalescence of consecutive droplets, (ii) generation of smaller droplets, which are detached from larger ones and picked up by others, and, the most crucial one, (iii) the wetting of the channel walls by the droplet liquid. In this case, residues located on the walls can be picked up by other droplets with which they coalesce \cite{ref:Kaminski2016}. Additionally, osmotic effects are always at stake and active agents may diffuse into the carrying phase, at least partially, causing uncontrolled variations of initially defined active dose per drop. Other issues are related to the mechanical design and the materials of the microfluidic device \cite{ref:Zhang2006, ref:Tiwari2020}. Miniaturizing of mechanical components inevitably leads to higher requirements regarding production accuracy, and traditional production processes need to be reconsidered or adapted. The choice of the channel materials itself is crucial as it may favour the previously mentioned unwanted wetting of the walls. Finally, and most critically, the high risk of clogging of the microchannels needs to be carefully considered \cite{ref:Hassanpourfard2016, ref:Dressaire2017}. In particular, when the transport and the encapsulation of microparticles or cells come into play, complex phenomena may appear at bottlenecks, deflections and at the liquid/solid interfaces, which lead to particle aggregation. Clog formations dramatically influence the performance of microfluidic devices and may instantaneously stop the production.

By taking these difficulties into account, some researchers have developed a totally different concept, which avoids most of these issues by design. This approach is often called in-air microfluidics (IAMF) \cite{ref:Planchette2018_ILASS, ref:Planchette2018, ref:Visser2018, ref:Kamperman2018}. The idea is to replace the channels of the microfluidic device by a continuous liquid jet (down to micrometer scale) and to manipulate and combine the components, for example droplets, in air. The concept does not require any carrying phase or channel and, consequently, clogging cannot occur. Moreover, thanks to the absence of viscous stresses at the channel walls, the throughput of in-air microfluidics can be orders of magnitude greater than those offered by classical microfluidic devices \cite{ref:Chiu2017, ref:Dressaire2017}. This also implies that, for processing comparable volumes, IAMF requires much less energy than standard chip-based technology. One of the most promising application of IAMF is the production of capsules and fibers with and without regular inclusions. The switch from capsules to fibers is rather versatile as it simply requires to move from a regularly fragmented jet to a stable one. In contrast, the state of the art of complex fiber production relies on different electrospinning methods, such as coaxial or emulsion electrospinning, which cannot be used to obtain capsules. Further, these techniques show applicability limits, especially when it comes to controlling the fiber geometry. Coaxial electrospinning is rather limited to fibers with core-shell structures. For its part, emulsion electrospinning enables to encapsulate independent inclusions. Yet, it does not offer the regularity needed in most applications as it is rather impossible to keep a constant diameter or to control the size and position of the inclusions \cite{ref:Qi2006, ref:Moghe2008, ref:Yarin2011, ref:Hu2014, ref:Sanchez2016, ref:Enizi2018}. 

Pioneering steps towards a successful solidification of the liquid structures generated by in-air microfluidics, and thus towards possible technical applications, have been done by Visser et al. \cite{ref:Visser2018} and Kamperman et al. \cite{ref:Kamperman2018} not long ago. They firstly demonstrated the wide potential of in-air microfluidics via the in-flight collision of a regular droplet stream and a jet, or via the collision of two jets, either manufacturing particles or simple fibers without inclusions. The solidification method of the structures they selected is the ionic crosslinking of an aqueous alginate solution via calcium ions. In addition to that, Jiang et al. \cite{ref:Jiang2021} very recently demonstrated that the production of particles or simple fibers without inclusions is also possible via UV-polymerization. Here, it is important to note that, to our knowledge, studies demonstrating the possibility to use in-air microfluidics for manufacturing fibers with regular inclusions are still missing. The uniqueness of our approach and of the resulting encapsulation geometry could be of advantage in many domains \cite{ref:patent}. Solidified fibers with regular inclusions can potentially be used for encapsulating cells into a protecting and nutrient environment, or for delivering drugs with individualized amount of active agents in a controlled manner. Moreover, while the throughput shown in the studies of Kamperman et al. \cite{ref:Kamperman2018}, Visser et al. \cite{ref:Visser2018} and Jiang et al. \cite{ref:Jiang2021} (up to $100\,ml/h$) is orders of magnitudes larger than the one obtained with classical microfluidic devices ($\approx O(1)\,\mu l/h$) \cite{ref:Lei2018}, higher values may be desirable. This point as well as low production costs are especially relevant to industrial applications, i.e. to the successful implementation of a concept into a profitable technology. In this study, we demonstrate how fibers with and without regular inclusions can be produced with a throughput an order of magnitude larger than the one achieved by Kamperman et al., Visser et al. and Jiang et al., namely in the range of $>1400\,ml/h$ (corresponds to $>17 000\,m$ fiber per $h$). We test seven different liquid combinations and obtain seven differently prepared types of fibers. In doing so, we do not only change the crosslinking agent (type of divalent cations), but also the relative amount of the various ingredients, including polyethylene glycol (PEG) and glycerol. These fibers are then characterized by measuring their Young's modulus and their elongation at break. In addition to that, the equilibration of the fibers under controlled laboratory conditions is considered, and the differences between equilibrated (or dry) and wet fibers are presented. 

This work is organized as follows. First, the production method is presented, including a description of the experimental set-up, or prototype, and the typical parameter values set in this study. Then, the composition of the solutions at stake, their respective properties, and the selected combinations used to produce the fibers are presented. Section \ref{chap7:fiber_characterization} deals with the characterization methods applied to the produced fibers and discusses the need of controlled equilibration as a prerequisite for reproducible measurements. The results, i.e. the properties of the fibers obtained under various conditions are compared and interpreted in section \ref{chap7:discussion}. Finally, the paper ends with the conclusions.

\section{Production method}
\label{chap7:exp_setup}

\subsection{Principle}
\begin{figure}[t!]
\centering
  \includegraphics[width=15cm]{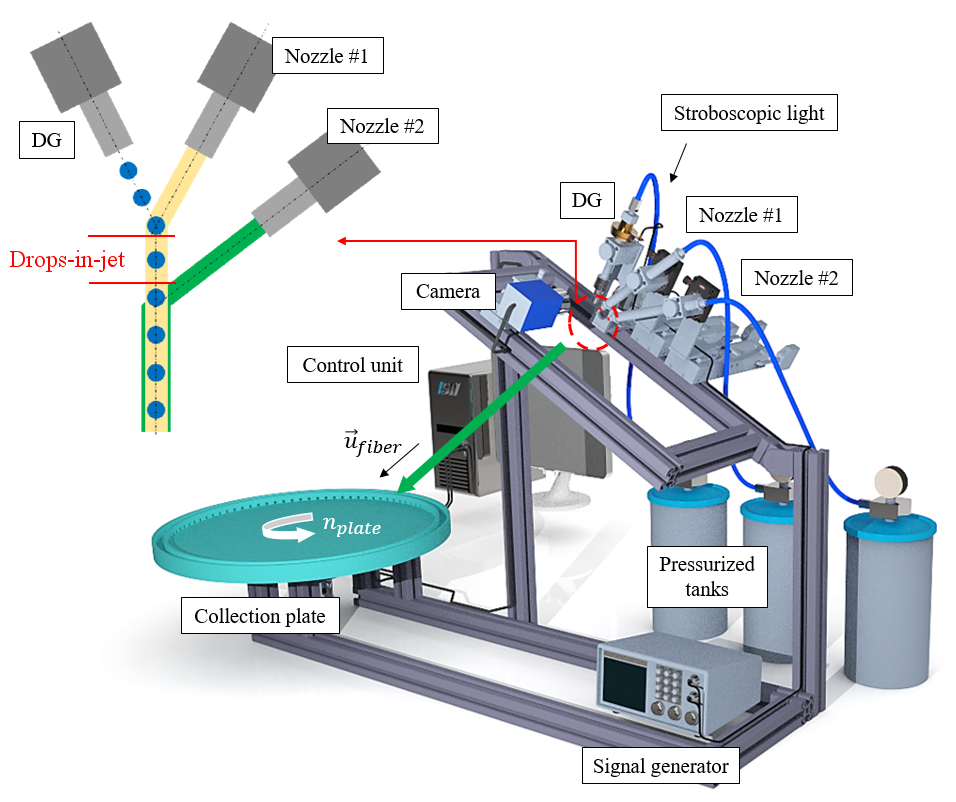}
 \caption{Schematic illustration of the experimental set-up (prototype) designed for the production of advanced fibers. Left: zoom into the two subsequent collisions respectively involving the droplets (blue) and the jet$_1$ (yellow); and the \textit{drops-in-jet} structure with the jet$_2$ (green). Right: overview of the various components enabling the controlled collisions.}
 \label{fig:7.1}
\end{figure}

The principle of our production technique relies on two subsequent steps: first, the generation of a controlled liquid structure called \textit{drops-in-jet} via the collisions of a droplet stream and a continuous liquid jet, followed, in a second step, by its solidification via a sol-gel transition. The whole process takes place in air where the resulting fibers are collected by a spinning disk. This principle and associated prototype are depicted in Fig. \ref{fig:7.1} while pictures taken during the process are shown later, in Fig. \ref{fig:7.2}. 

More precisely, the first jet (jet$_1$, yellow), made of an aqueous alginate solution, constitutes the basis of the fiber. The regularly impacting droplets (blue) are in turn building the regular inclusions of the future fiber. Thus, it is essential that the droplets get regularly embedded in the first jet, and that neither the droplets nor the jet fragment. The collision parameters are therefore adjusted to stably provide the desired \textit{drops-in-jet} regime, see red section in Fig. \ref{fig:7.1} and \ref{fig:7.2}. Other collision outcomes and associated fragmentation processes of the drop and jet, have been widely studied by our group and are presented in details elsewhere \cite{ref:Baumgartner2020, ref:Baumgartner2020_PRF, ref:Baumgartner2022_JFM}. In order to achieve an in-flight sol-gel transition of the generated \textit{drops-in-jet} structure, a second jet (jet$_2$, green) is needed, which immediately engulfs the previously formed compound (droplets + jet$_1$). This second step enables the quick solidification of the regular liquid structure into a regular fiber. By temporarily stopping the drop stream, it is therefore possible to temporarily suppress the inclusions in the fiber. Similarly, different droplet streams can be used to produce inclusions of different sizes, with different spacings or to encapsulate different actives. With or without inclusions, the fast encapsulation is essential to achieve uniform solidification of the alginate-based jet. Therefore, the second jet contains, beside the divalent cations which initiate the crosslinking reaction, a high ethanol content which promotes the encapsulation via the reduction of its surface tension. Note that the surface tension reduction could also be obtained with other chemicals, for example using surfactants. Details about the liquids and their preparation can be found in section \ref{chap7:materials}.

\subsection{Prototype}

The prototype enabling the fiber production comprises three pressurized tanks for the separate supply of the drop and jet liquids. The air pressure inside each of the three tanks and, relating thereto, the flow rates of the three liquids are independently adjusted using the OB1 MK3+ pressure controller from Elveflow (Paris, France). The droplet stream is generated with a droplet generator (DG) \cite{ref:Brenn1996} used with an orifice of diameter $D_{orifice}=100 \mu m$, which is connected to an external signal generator. The two jets are created using two nozzles (nozzle \#1 and nozzle \#2) with diameters $D_{nozzle,1}=340 \mu m$ and $D_{nozzle,2}=230 \mu m$, respectively. The droplet generator is fixed, while the two nozzles are mounted on microtraverses for an accurate adjustment enabling, among others, the suppression of any off-plane eccentricity. 
Moreover, we use a camera with a resolution close to $4 \mu m /px$ to get detailed collision pictures in the collision plane. The collisions are illuminated from the backside using a stroboscopic light, which is connected to the same signal generator as the droplet generator in order to produce standing pictures. The frequency is set to $f_d=15500 Hz$ for all tested liquid combinations. It is important to note that the continuous monitoring of the off-plane eccentricity and of the preset collision parameters during the production process is supported by a software application implying the control unit, the camera and the automated (motorized) microtraverses. Finally, the generated fibers must be collected without inducing too much stress and strain. At this early production stage, the fibers are made of wet hydrogels, which are fragile and easily stretchable, requiring a soft handling. The collection system therefore consists of a horizontal plate mounted on  a brushless  motor, which is connected to the control unit. The speed of the spinning plate $n_{plate}$ is controlled to match the one of the fiber at its impact point $\Vec{u}_{fiber}$. After the fibers are collected, they are let to dry under room conditions before further process steps follow. Details can be found in section \ref{fiber_drying}.

\subsection{Parameter space}

\begin{figure}[b]
\centering
  \includegraphics[width=\textwidth]{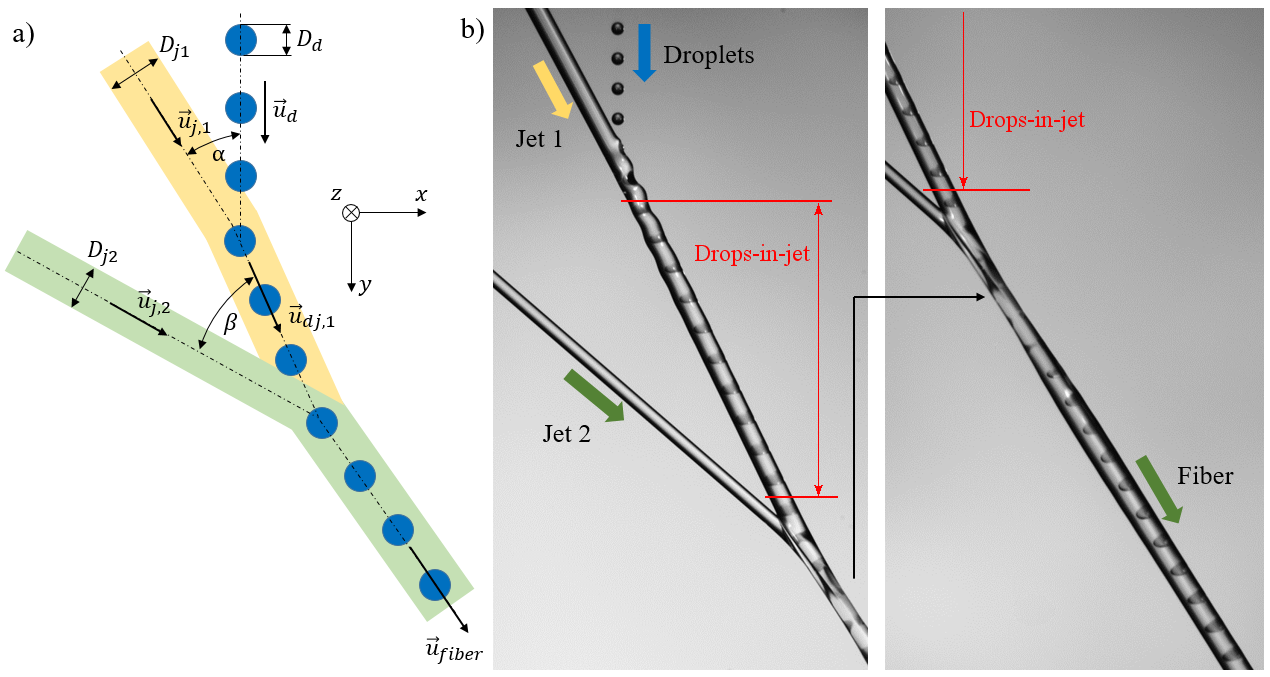}
 \caption{(a) Kinetic and geometric parameters, which are relevant for the production of fibers with regular inclusions. (b) Pictures of the collisions observed in the collision plane. The collisions between the droplets and the jet$_1$ enable the formation of the liquid \textit{drops-in-jet} structure, which is subsequently solidifies thanks to the supply of cations ensured by the collision with the jet$_2$. }
 \label{fig:7.2}
\end{figure}
All relevant kinetic and geometric parameters are shown in Fig. \ref{fig:7.2}(a) and are extracted from the collision pictures, see Fig. \ref{fig:7.2}(b). These parameters do not change much when testing the different liquid combinations. The diameters of the droplets and the jets are typically in the range of $D_d=150 \pm 10 \mu m$, $D_{j,1}=320 \pm 5 \mu m$ and $D_{j,2}=213 \pm 5 \mu m$, respectively. Here, the subscripts $d$, $j,1$ and $j,2$ stand for the droplet, the first and the second jet, respectively. The two collision angles, $\alpha$ and $\beta$, are set to $25.5 \pm 1.5^{\circ}$ and $28 \pm 2^{\circ}$, respectively. The absolute value of the droplet velocity, given by $\Vec{u}_d=\Vec{L}_df_d$, is $6.7 \pm 0.8 ms^{-1}$, where $\Vec{L}_d$ is the distance between two consecutive droplets. The two flow-rate equivalent jet velocities $|\Vec{u}_{j,1}|$ and $|\Vec{u}_{j,2}|$ are measured and both are found to be $5.0 \pm 0.2 ms^{-1}$. Further, the velocity of the generated structure $\Vec{u}_{fiber}$, is estimated via a momentum balance. It is found to remain in the range of $5.0 \pm 0.2 ms^{-1}$, showing that the contribution of the droplets to the overall momentum is almost negligible. The relative impact velocities $\Vec{U}_1=\Vec{u}_{d}-\Vec{u}_{j,1}$, between droplets and jet$_1$, and $\Vec{U}_2=\Vec{u}_{j,2}-\Vec{u}_{dj,1}$, between jet$_2$ and droplets+jet$_1$, are calculated providing values of $3.0 \pm 0.5 ms^{-1}$ and $1.9 \pm 0.2 ms^{-1}$, respectively. The impact Weber numbers of the subsequent collisions, i.e. the one of the droplets with respect to jet$_1$, $We_d=\rho_d D_d U_1^2/\sigma_d$, and the one of the \textit{drops-in-jet} with respect to jet$_2$, $We_{j,2}=\rho_{j,2} D_{j,2} U_2^2/\sigma_{j,2}$, remain smaller than $35$ for all tested liquid combinations. These small values of $We_d$, in combination with $L_{j,1}/D_{j,1}=u_{j,1}/(D_{j,1}f_d)\approx1$ obtained for all sets of experiments, ensure that the \textit{drops-in-jet} structure is stably established for all investigated collisions \cite{ref:Baumgartner2020, ref:Baumgartner2020_PRF}.  

\section{Materials and liquid properties} 
\label{chap7:materials}
\subsection{Investigated combinations} 

Several studies on alginate-based fibers show that their mechanical properties are strongly affected by the manufacturing conditions, the molecular weight and the structure of the sodium alginate, the type of cross-linking agent, possible additives and their concentrations in the solutions, as well as by the time of contact between the cross-linking agent and the alginate \cite{ref:Niekraszewicz2009, ref:Lee2012}. 

Thoroughly testing all these effects would go far beyond the scope of this work. Instead, we decide to focus on the most relevant parameters, which results in the investigation of seven different types of fibers, whose production conditions are listed in Table \ref{tab1:liquid_combination}. The precise composition of each solution and corresponding properties are given in the next section. Using preliminary results (not shown), we first define a reference fiber (\textit{Fiber ref} in Table \ref{tab1:liquid_combination}). This fiber contains inclusions formed by droplets and is produced with calcium ions as cross-linking agent, with polyethylene glycol (PEG) as plasticizer and without any subsequent post-treatment, i.e. relying only on the sol-gel transition happening in air.

To test the influence of the contact time between the cross-linking agent and sodium alginate, a fiber similar to the reference fiber is produced and placed, immediately after collection, in a liquid bath. The bath contains the same solution as the jet$_2$. The residence time of the fibers in the bath is fixed to 5, 10, and 30 minutes. The resulting fibers are named \textit{Fiber bath 5min}, \textit{10min} and \textit{30min}, respectively, see Table \ref{tab1:liquid_combination}.

Next, a fiber without inclusions is produced (\textit{Fiber w/o droplets} in Table \ref{tab1:liquid_combination}). Practically, the same conditions as for the reference fiber are used but the droplet stream is switched off, leaving only the jet$_1$ and jet$_2$ interact.

In order to evaluate the influence of a plasticizer on the mechanical properties of the hydrogel, we prepare for the jet$_1$, a sodium alginate solution without PEG. Different studies show that, PEG blended with sodium alginate can modify the elasticity of the resulted fiber, which further affects its Young' modulus as well as the elongation at break \cite{ref:Wang2007, ref:Dey2012}. The corresponding fiber is called \textit{Fiber w/o PEG} in Table \ref{tab1:liquid_combination}.

Finally, the influence of the crosslinking agent is investigated. Thus, we replace the widely used calcium chloride, which provides calcium cations, by strontium chloride, which supplies the strontium cations at similar concentration. Note that whatever the crosslinking agent, it is always present both in jet$_2$ and in the droplets. Thus, to obtain this last fiber, called \textit{Fiber SrCl$_2$} in Table \ref{tab1:liquid_combination}, we change both the droplet liquid and the one of jet$_2$.

\begin{table}[t]
\line(1,0){510}
\centering
\caption{Different fiber types tested in this study and the corresponding conditions used for their production. Liquid combination and composition for the droplets, the jet$_1$, the jet$_2$ and the liquid bath (if any). [x = yes; - = no]}
\begin{ruledtabular}
\setlength{\tabcolsep}{1.15em}
\begin{tabular}{l|c|c|c|c|c|c|c|}
  & \multicolumn{7}{c}{Liquids} \\ [3 pt]
  &  \multicolumn{2}{c}{Droplets} & \multicolumn{2}{c}{Jet$_1$} & \multicolumn{2}{c}{Jet$_2$} & Bath \\ [3 pt]
  & G5  & G5  & \,\,\,\,\, Alg \,\,\,\,\, & Alg  & EtOH  & EtOH  & EtOH   \\ 
    &  + CaCl$_2$ &  + SrCl$_2$ &  &  w/o PEG &  + CaCl$_2$ &  + SrCl$_2$ &  + CaCl$_2$  \\ [3 pt] \hline 
    & & & & & & & \\ [-5 pt]
       1. Fiber ref  & x & - & x & - & x & - & -  \\
       2. Fiber bath 5min  & x & - & x & - & x & - & x  \\
       3. Fiber bath 10min  & x & - & x & - & x & - & x  \\
       4. Fiber bath 30min  & x & - & x & - & x & - & x \\
       5. Fiber w/o droplets \, & - & - & x & - & x & - & - \\
       6. Fiber w/o PEG & x & - & - & x & x & - & - \\ 
       7. Fiber SrCl$_2$  & - & x & x & - & - & x & - \\ [3 pt]
\end{tabular}
\end{ruledtabular}
\label{tab1:liquid_combination}
\end{table}

\subsection{Solution composition and properties}

\begin{table}[b]
\line(1,0){510}
\centering
\caption{Properties of the liquids used in this study. All measurements were carried out at ambient conditions of $T= 23 \pm 1^{\circ}$C and $ RH = 40 \pm 3\%$.}
\begin{ruledtabular}
\setlength{\tabcolsep}{1em}
\begin{tabular}{lccc}
Liquid  & Density  & Dynamic viscosity  &  Surface tension \\
        &  $\rho$ (g\,dm$^{-3}$) &  $\mu$ (mPa\,s) & $\sigma$ (mN\,m$^{-1}$) \\ [3 pt]
    \hline \\ [-5 pt]
       G5+CaCl2   & 1187.5 & 7.9  & 65.2 \\
       G5+SrCl2   & 1210.1 & 5.7  & 65.8  \\
       Alg  & 997.4  & 53.1  & 45.5  \\
       Alg w/o PEG & 996.6 & 39.3 & 45.5 \\
       EtOH+CaCl2 & 981.6 & 3.3 & 26.6  \\
       EtOH+SrCl2 & 1035.1 & 3.2 & 27.3\\ [3 pt]
\end{tabular}
\end{ruledtabular}
\label{tab2:liquid_properties}
\end{table}

\begin{itemize}
    \item Jet$_1$ or alginate solution

In this study, the molecular weight of the sodium alginate (low molecular weight, low viscosity) and its concentration in the liquid base solution (jet$_1$) are kept constant for all tested liquid combinations. Note that, the exact molecular weight of the sodium alginate cannot be provided by the supplier. To avoid uncontrolled variations, the same batch is used throughout the study. The selected sodium alginate, purchased from Sigma Aldrich, USA, as low viscosity alginic acid sodium salt from brown algae with 39\% G-block and 61\% M-block, allows us to use reasonably high concentrations while keeping the solution dynamic viscosity moderate, see Table \ref{tab2:liquid_properties}. Hence, the aqueous alginate solutions can be jetted with sufficiently high jet velocities ($\Vec{u}_{j,1}$), while the pressure drops occurring in the nozzle and in the whole liquid supply system remain acceptable. The best composition has been specified via extensive preliminary studies using different types of sodium alginates with different concentrations. 

For the reference fiber, the alginate solution, \textit{Alg}, further contains polyethylene glycol 20000 ($M_w\approx 20000 g/mol$ from Carl Roth GmbH, Germany), both being dissolved in a mixture of distilled water and ethanol (from Carl Roth GmbH, Germany). The mass fractions of the four components: sodium alginate, PEG, water and EtOH are $3.5 \%$, $1.5 \%$, $85 \%$ and $10 \%$, respectively.

The second alginate solution used for jet$_1$, \textit{Alg w/o PEG}, is similar to \textit{Alg} except that no PEG is added. Thus, the mass fractions of the three components: sodium alginate, water and EtOH are $3.5 \%$, $86.5 \%$ and $10 \%$, respectively. 

In both cases, the addition of ethanol has two effects. First, it decreases the surface tension of the jet liquid promoting the encapsulation of the droplets. Visser et al. \cite{ref:Visser2018} show that a surface tension difference $\Delta \sigma \geq 10 \, mN/m$ is sufficient to achieve encapsulation. Second, it is expected to improve the mechanical properties of the hydrogel network, as previously reported by Hermansson et al. \cite{ref:Hermansson2016}. 

 \item Droplets or G5 solution
 
It is important to mention that, the crosslinking agents are not only added to the liquid of the second jet (jet$_2$), but also to the one of the droplets. The reason for this is two-fold. First, it supports the in-flight solidification of the alginate solution making it sufficiently fast to collect the fibers. In practice, the distance between the collision point and the collection plate is approximately $1m$ which leads for typical fiber velocities $u_{fiber}\approx5ms^{-1}$ to an available solidification time of only $200ms$. Even though the gelation transition of sodium alginate dissolved in water via ionic crosslinking is known to be fast compared to other gelation processes \cite{ref:Sun2013}, it is preferable to enhance this transition by initiating it from both the inside (via the droplets) and the outside (via jet$_2$) of the main jet (jet$_1$). Secondly, the divalent cations added to the droplets help to fix the shape and position of the droplets (inclusions) inside the jet made of alginate solution as regularly as possible. Indeed, the three liquids, which are brought into contact, are miscible. To limit the diffusion of the droplet liquid into the jet$_1$ liquid, and reciprocally, it is advantageous to trigger additional and quasi-instantaneous gelation upon the contact between them.

For the drop liquids (\textit{G5+CaCl2} and \textit{G5+SrCl2}), we use an aqueous glycerol solution with a mass fraction of glycerol ($\geq 98 \%$,
Carl Roth GmbH, Germany) in distilled water (Kerndl GmbH, Germany) of 50\%. In order to test the influence of the crosslinking agent, we add calcium chloride CaCl$_2$ (Carl Roth GmbH, Germany) and strontium chloride SrCl$_2$ (in the form of strontium chloride hexahydrate SrCl$_2\cdot$6H$_2$0, $\geq 99 \%$ p.a., Carl Roth GmbH, Germany) both with the same concentration of $1\,mol/l$ to the droplet liquid G5. Keep in mind that, 1 mole of strontium chloride hexahydrate contains the same number of strontium atoms than 1 mole of pure strontium chloride, which, in turn, is equal to the number of calcium atoms in 1 mole of calcium chloride. Thus, the concentrations are comparable. The droplet liquids are dyed with Rhodamin B (C.I.45170, Merck, USA).

\item Jet$_2$ or cross-linking solution

The two aqueous ethanol solutions of jet$_2$ (\textit{EtOH+CaCl2} and \textit{EtOH+SrCl2}) are, used to probe the influence of the different divalent cations as cross-linking agent. Calcium chloride CaCl$_2$ and strontium chloride SrCl$_2$ are added with the same concentration as for the droplet liquids ($1\,mol/l$). The aqueous ethanol solution, which is used for both liquids, is a mixture of distilled water and ethanol with a mass ratio of 50\%:50\%. This high concentration of ethanol is used to strongly increase the surface tension difference $\Delta \sigma$ between jet$_1$ and jet$_2$, which leads to a quick coating of jet$_1$ by jet$_2$ and thus promotes a uniform solidification. As mentioned already, depending on the need of the application, and especially if cells are used, other chemicals could be used to lower the surface tension. 

\item Bath

When a liquid bath is used, its composition is the same as the one of the jet$_2$. Thus, in the presently tested configurations, it consists of a water and ethanol mixture at 50:50(w:w), in which CaCl$_2$ is dissolved at a concentration of $1\,mol/l$. It is referred to as \textit{EtOH+CaCl2}.

\end{itemize}

For completeness, the values of the surface tension $\sigma$, the density $\rho$, and the dynamic viscosity $\mu$ of all previously listed liquids are shown in Table \ref{tab2:liquid_properties}. The density is measured by weighing an exact volume of $100 ml$ (graduated flask and analytical scale), the viscosity is determined with a glass capillary viscometer, and the surface tension is measured with the pendant drop method.

\section{Fiber characterization}
\label{chap7:fiber_characterization}
In this work, the regularity of the fibers is first optically assessed. Then, the fibers are characterized by their mechanical properties, and more precisely, by their elongation at break $\epsilon$ and Young's modulus $E$. To guarantee reproducibility, all measurements are carried out after the fibers have been pre-conditioned. This step consists in storing the fibers under controlled conditions (temperature and humidity) for a minimum period of time until they equilibrate with these conditions. This conditioning step is justified by the evolution of $\epsilon$ and $E$ observed right after fiber collection and is  presented in this section, after the measurement methods. 

\subsection{Optical control of regularity }

\begin{figure}[h]
\centering
 \includegraphics[width=8cm]{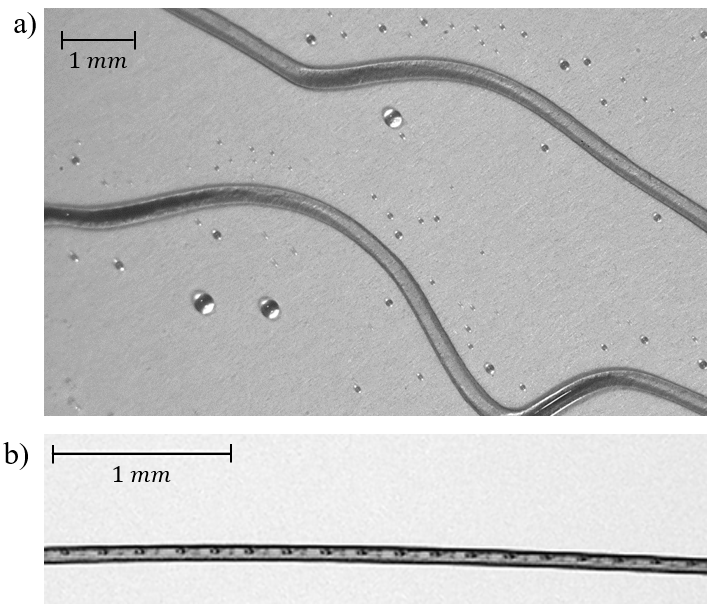}
  \caption{Recorded picture of (a) \textit{Fibers w/o droplets} in wet state showing the uniformity of the fibers immediately after collection and (b) the reference fiber type \textit{Fiber ref} with regular inclusions in dry state.}
   \label{fig:7.6}
\end{figure}

One of the key-point of our method compared to classical electrospinning approaches is the regularity of the produced fibers. The latter have a cylindrical shape with a constant diameter. Further, when inclusions are present, they have a well-defined size and spacing period. To control the regularity of the fibers, we image them and measure their diameter. We thoroughly and consistently obtained variations of less than $5\%$ and therefore do not present or discuss the data further. Fig. \ref{fig:7.6} illustrates this aspect with pictures of (a) \textit{Fibers w/o droplets} showing the uniformity of the generated fibers immediately after the collection, and (b) the fiber reference type \textit{Fiber ref} with regular inclusions in equilibrated state.

\subsection{Measurement methods: Elongation at break and Young modulus }
\begin{figure}[b]
\centering
    \includegraphics[width=14.5cm]{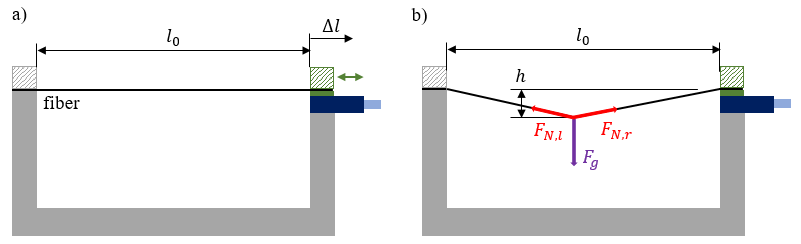}
  \caption{Measurement principle and sketch of the device used for the determination of the elongation at break $\epsilon$ (a), and Young's modulus $E$ (b).}
  \label{fig:7.3}
\end{figure}

To obtain $\epsilon$ and $E$, we use the same measurement device, which was designed and built in-house. Schematic illustrations are shown in Fig. \ref{fig:7.3}. The device consists of a base (grey, fully coloured), which is fixed, and a one-axis micro-traverse (blue, fully coloured) mounted on the base. The cross-hatched areas indicate the two parts, which are used to clamp the fibers. Further, the green coloured elements, attached to the micro-traverse, can be moved by operating it. It is important to note that, all edges, with which the fibers may get into contact, are rounded off to eliminate irregularities, which could damage the fibers. For the same reason, surfaces clamping the fibers are well polished. 

For the measurement of the elongation at break $\epsilon=l_b/l_0$, the fiber is carefully clamped in the measurement device, whereby slacking must be avoided, see Fig. \ref{fig:7.3}(a). At this point, a camera is used to measure the initial length of the fiber $l_0$. After that, the samples are quasi-statically elongated with the help of the micro-traverse until the fiber breaks. The whole elongation process is recorded as video for an accurate determination of the breaking point. The breaking length of the fiber $l_b=l_0+ \Delta l$ is measured with the help of the last recorded frame showing an intact fiber.

The determination of the Young's modulus $E=\Delta \sigma/ \Delta \epsilon$ is performed with the same device, see Fig. \ref{fig:7.3}(b). This parameter quantifies the slope of the stress-strain diagram of a certain elastic material. Here, $\sigma$ stands for the normal stress in the fiber. At the beginning, the fiber is again carefully clamped in the measurement device, whereby slacking must be avoided, and a picture is taken to estimate $l_0$. In the next step, the first weight $m_1$ is placed in the middle of the fiber (at $ \approx l_0/2$) and another picture is recorded. The added weight leads to an elongation of the fiber $\epsilon_1$, while the fiber is subjected to a force $F_{g,1}=m_1g$. The elongation, corresponding to $m_1$, can be calculated with  $\epsilon_1=(1+4h_1^2/l_0^2)^{1/2}-1$. The weight force $F_{g,1}$ is balanced by tensions $F_{N,l}$ and $F_{N,r}$ on each side of the fiber. In the case of symmetry, $F_{N,l}=F_{N,r}=F_1$ and the induced stress can be calculated with $\sigma_1=F_1/A$, where $A=D^2 \pi/4$ defines the cross section of the fiber and $D$ is the fiber diameter. This procedure is carried out by using five weights in total with an increasing mass from $m_1$ to $m_5$, varying between $0.0312\, g$ and $0.5017\, g$. This leads to a stress-strain diagram including five data points, for each fiber. By fitting these points with a linear function, the Young's modulus $E$ can be estimated. To ensure that the stretching of the fiber is in the range of an elastic deformation, pictures are taken after each weight is removed from the fiber. These pictures are used to control the fully elastic relaxation of the fiber to its initial stage. Note that, the linear fitting function of the stress-strain relation usually shows a small positive intercept, i.e. a non-zero stress at $l_0$. This can be attributed to some pre-strain of the fiber. In order to compensate this slight offset, $l_0$ is corrected and, consequently, the respective measured strain values get slightly modified. The resulting Young modulus values, however, do not change significantly when implementing this correction since it represents a variation of less than $0.5\%$.

\subsection{Fiber pre-conditioning} \label{fiber_drying}

After the different types of fibers are produced, similar temperature and humidity conditions are applied in order to get reliable and comparable measurements of their characteristics. Indeed, immediately after the fibers are collected, their hydration level is quite high and their wet state is source of difficulties for reproducible characterization. This aspect is known and different studies show that the mechanical properties strongly deviate when measuring the same fiber or gel in dry and wet state \cite{ref:Zeugolis2009, ref:McNamara2019, ref:Wei1992, ref:Li2016_fibers}.

To characterize this variability, we first follow the evolution of the reference fiber (\textit{Fiber ref}) properties during the first instants after its collection, i.e. in its wet state. The information obtained in this way motivate the pre-conditioning of the fibers. The duration of preconditioning is determined in a second step, by measuring the fiber diameter, as explained at the end of this section. In all experiments, the conditions in which the fibers are left to equilibrate correspond to the ambient conditions of our laboratory, which are controlled. During the whole study, the temperature is found to be $T=23 \pm 1 ^{\circ}C$ and the relative humidity $RH=40 \pm 3 \%$.

\begin{figure}[b]
\centering
    \includegraphics[width=\textwidth]{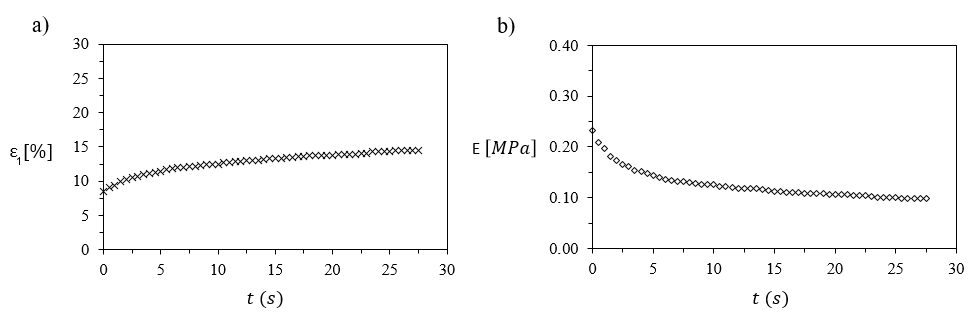}
  \caption{(a) The elongation with a fixed weight (approximately constant stress) $\epsilon_1$ and (b) the pseudo Young's modulus $E$ as a function of time of the fiber reference type \textit{Fiber ref} in wet state, i.e. immediately after the production and collection. }
  \label{fig:7.4}
\end{figure}

To follow the evolution of the fiber properties during its first instants, the wet fiber is placed immediately after collection, in the measurement device. To enable a better control of the fiber "age", the measurements are preformed by a team of 2 operators. Further, the protocols previously described are slightly modified. More precisely, a single freshly produced sample is used in combination with a single weight and the temporal evolution of the fiber elongation, $\epsilon_1$, is followed taking the instant when the weight has been added as time origin $t=0s$. Note that $\epsilon_1$ is  not the fiber elongation  measured just before it breaks, but the one obtained for a fixed weight, which can - at first order - be considered as a fixed stress.  These data are further used to determine the evolution of the \textit{pseudo} Young's modulus. Here, the term pseudo is used to indicate that  the fiber deformation is not elastic and that per definition, the slope of the strain-stress curve does not strictly correspond to the Young's modulus. The results are plotted in the form of $\epsilon_1(t)$ and $E(t)$ in Figs. \ref{fig:7.4}(a) and (b), respectively. Shortly after the weight has been added, the fiber continuously deforms and, as a result, the elongation $\epsilon_1$, with respect to a constant stress, and $E$ are not fixed anymore. This can be attributed to the fact that (i) the fiber behaves like a hydrogel, more specifically, like a wet viscoelastic hydrogel \cite{ref:Mancini1999, ref:Chaudhuri2017} and not like a solid network and that (ii) the residual water, which is stored in the wet fiber, evaporates during the measurement procedure leading to a change of the fiber hydration level. This continues until an equilibrium state between the fiber and the surrounding environment is reached. These observations make it almost impossible to compare the different types of fibers in wet state in a meaningful manner. Thus, we decide to let the fibers fully equilibrate with the controlled laboratory conditions, before further characterizations are done.

\begin{figure}[t]
\centering
    \includegraphics[width=14cm]{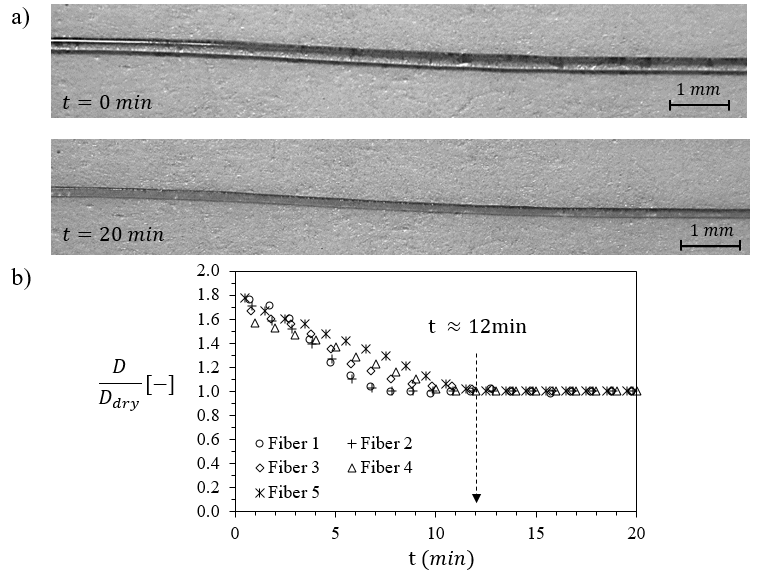}
  \caption{(a) Recorded pictures of a  \textit{Fiber w/o droplets} in wet ($t=0\,min$) and dry state (after $t_{dry}>12\,min$, at $T=23 \pm 1^{\circ}C$ and $RH=40 \pm 3\%$); (b) Diameters $D$ of five fiber samples immediately after collection, normalized by their diameters in dry state $D_{dry}$, as a function of time. Used to specify the drying process of the generated fibers.}
  \label{fig:7.5}
  \vspace{0.2cm}
\end{figure} 

In order to define the minimum drying or pre-conditioning period at ambient conditions, we produce five samples of the type \textit{Fiber w/o droplets} (see Table \ref{tab1:liquid_combination}). These samples are placed directly after collection, on a water-repellent substrate (solid substrate covered with a polypropylene foil) and their diameters are measured at regular intervals with the help of a camera. Fig. \ref{fig:7.5}(a) shows an example of the \textit{Fiber w/o droplets} (Fiber 3, empty diamonds) in wet ($t=0\,min$) and dry state ($t_{dry}>12\,min$). The diameters of the five samples $D$, normalized by their diameters in dry state $D_{dry}$, are plotted as a function of time in Fig. \ref{fig:7.5}(b). The equilibration rate is similar for all five samples before reaching the steady state plateau, which happens after $t_{dry}\approx 12 \, min$. From there on, the fiber diameter does not further change, which supports the assumption that the equilibrium state is reached.  Note that, the time period of $t_{dry}\approx 12 \, min$ is defined as the minimum fiber equilibrating time after collection for all types of fibers produced within the scope of this work. In practice, we always characterize fibers after they equilibrate for more than $60 \, min$.

\section{Results and discussion}
\label{chap7:discussion}

In this section we focus on the Young's modulus $E$ and the elongation at break $\epsilon$ of the different fiber types discussed in section \ref{chap7:materials}. The characterization procedures are explained in section \ref{chap7:fiber_characterization} and all measurements are carried out at ambient conditions with fibers totally equilibrated for at least $12\, min$ ($> t_{dry}$). Practically, the fibers are produced, collected, and hanged to equilibrate in air (laboratory conditions) leading to a uniform dehydration of the samples. 

\subsection{Young's modulus}
\begin{figure}[t!]
\centering    \includegraphics[width=15cm]{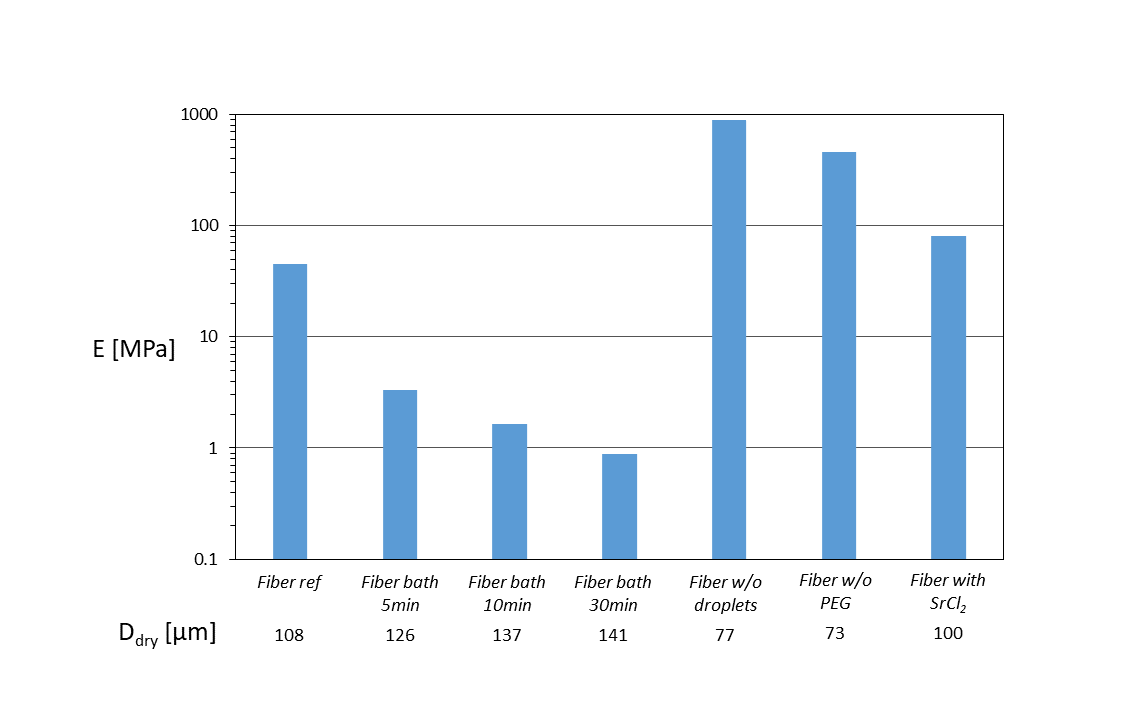}
  \caption{The Young's modulus $E$ of the seven different types of fibers characterized in this study and their diameters $D_{dry}$.}
  \label{fig:7.7}
\end{figure}

The Young's modulus of the seven differently prepared fibers and their diameters are shown in Fig. \ref{fig:7.7}. Here, at least five samples of each type are characterized and only the highest value of $E$ and the corresponding diameter $D_{dry}$ is represented in the figure. The reason for this is that, the $E$-values of the samples related to one fiber type can deviate. The reasons are probably multiple and cannot be definitively identified. For example, and despite images showing very regular structures, one cannot totally exclude microscopic imperfections in these structures. Other homogeneities, for example of the equilibrating or cross-linking processes, could also cause these deviations. Despite careful adjustment of the collecting plate velocity with the one of the fibers, small differences cannot be avoided, which may stretch the fibers and thus modify their mechanical properties. Similarly, damages produced while placing the fragile fiber in the measurement device, especially at the clamping sites, are expected to affect the results. These damages are limited by the gentle handling of the fibers and the smooth surface of the clamping parts but cannot be totally eliminated. Finally, we identify the hanging of the weights as another potential source of very local, yet critical damages. The latter are indeed not relevant to our study but could constitute a large source of noise. To reduce this noise, we have chosen to keep the highest measured value of $E$ for each type of fiber. In this way, we expect the measurement to represent the less damaged fiber, and thus the most relevant value for comparison. Note that the hypothesis of local damages - either at clamping or by hanging the weights - is supported by the comparable low data dispersion obtained for elongation at break. For measuring the elongation at break, no weight is used and only fibers which break away from the two clamping points are analyzed. Thus, it also justifies, a posteriori, our choice of presenting the highest value of $E$ for each fiber type.

The Young's modulus of the reference fiber (with regular inclusions and without calcium chloride bath) $E_{Fiber\,ref}$ is found to be about $45 MPa$ in this work. It is important to note that, a cross-study comparison of $E$ obtained for different alginate-based fibers is extremely difficult, due the large amount of influencing factors. The fiber preparation, additives, molecular weights, cross-linking time, hydration level, etc. strongly affect the mechanical properties, as already demonstrated. The Young's modulus of alginate-based fibers, can vary of several order of magnitudes, for example, between $0.2\, MPa$ \cite{ref:Cuadros2012} if produced with a classical microfluidic device, and $6\,GPa$ \cite{ref:Sibaja2015}, if obtained by a wet spinning technique. 
Other factors, which may lead to large differences in determined values of the Young's modulus, may be the measurement device/technique itself, as well as the fact that the obtained stress strain relation of the fiber may not always be thoroughly linear. In the work by Cuadros et al. \cite{ref:Cuadros2012}, who used a Universal Texture Analyser TA.XT2i (Godalming, Surrey, UK) in tension mode, the stress strain relation of alginate fibers is clearly linear until the tested fiber fractures. Note that in our device, we only obtain linear strain stress curves, except for the small deviations at origin, which are caused by a small pre-strain of the fiber when fixing it and before loading it. Other studies, such as those of, Zhang et al. \cite{ref:Zhang2020} and McNamara et al. \cite{ref:McNamara2019}, who used a tensile tester XQ-1C (Shanghai, China) and an Instron Universal Testing machine model 5569 (Canton, MA, USA), respectively, show partly and even strongly non-linear strain stress curves. This, of course, questions the relevance of the deduced Young modulus. Thus, quantitative comparison of the Young's modulus is only made for the results obtained within this study, i.e. for fibers produced with the same method and characterized with the same device and protocol. Discussion including results from other groups remains at a qualitative level.

\begin{figure}[b]
\centering
    \includegraphics[width=10cm]{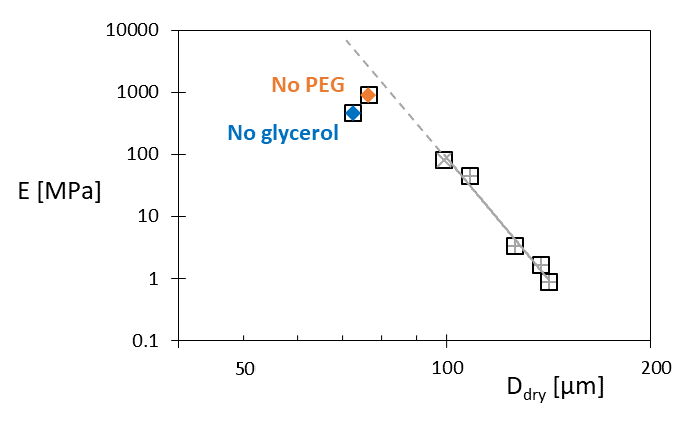}
  \caption{ Young's modulus, $E$, as a function of the diameter of the equilibrated fiber, $D_{dry}$. Same data as in Fig. \ref{fig:7.7} (empty squares); the fibers without droplet/PEG are marked with blue/orange diamonds; grey $+$/$\times$ indicates the presence of $Ca^{2+}/Sr^{2+}$. The continuous line shows the correlation given by $E=8~10^{27} {D_{dry}}^{-13}$ and the dashed line enables to visualize deviations observed in the absence of plasticizer.}
  \label{fig:7.7n}
\end{figure}

Placing the reference fiber \textit{Fiber ref}, just after collection from the spinning disk, in a liquid bath containing calcium chloride $CaCl_2$ during 5, 10 and 30 min, significantly decreases the Young's modulus, while it increases the fiber diameter. The Young modulus is reduced by as much as a decade if the fiber is immersed 5 minutes in a CaCl$_2$ bath. This reduction continues if the fiber is left longer in the bath. Yet, the effects seem to saturate. After 30 minutes and more (data not shown), the Young modulus remains close to 1MPa. Thus, averaged over the first 5 minutes and over the last 20 minutes, the decrease rate is found to be approximately $8 MPa / min$ and $0.04 MPa / min$, respectively. Similarly, the effects on the diameter are more important during the first 5 minutes. During this period, $D_{dry}$ increases of $17\%$, thus with an average rate of more than $3\% / min$ while during the last 20 minutes, the diameter changes only by less than $3\%$, which corresponds to an average rate of less than $0.3\% / min$.
The evolution of $D_{dry}$ can be attributed to the hygroscopic behaviour of calcium chloride \cite{ref:Vainio2019, ref:Vrana2014} in combination with the different retention times in the liquid bath. The longer the retention time, the more divalent calcium cations diffuse into the alginate network. Thanks to the higher calcium content and its hygroscopic behaviour, more water is present and the hydration level of the fiber remains high, even if the fiber is left to equilibrate with room conditions for several hours. This property is called water retention rate of sodium alginate fibers \cite{ref:Wang2019}. The large diameters simply correspond to higher content of residual water in the fibers. It is confirmed by the comparison of the lineic weight of the fibers. After drying, the reference fiber \textit{Fiber ref} has a lineic weight of $\approx 10\, \mu g/m$, while the one of the fibers, which was first placed in a liquid bath for 30 min, increases up to $\approx 27\, \mu g/m$. Yet, after some time in the bath, the adsorption of calcium cations reaches its maximum and saturates, explaining why the magnitude of these effects are stabilizing after 30 minutes. To understand the evolution of the Young modulus, it is important to note that the "hydrated" fibers rather behave like hydrogels and not like solid-like networks. Consequently, the increase of the fiber diameter is concurrent with the strong decrease of the Young's modulus down to less than $1\, MPa$. McNamara et al. \cite{ref:McNamara2019} also found that wet alginate fibers have a much weaker Young modulus than their dry counterparts. They also attribute this to a plasticizing effect of water in biopolymer fibers \cite{ref:Harper2013}. In general, different studies also observed that wet fibers are more fragile and plastic than the same fibers characterized in dry state \cite{ref:Wei1992, ref:Li2016_fibers}. This point is further discussed at the end of this section, where Fig. \ref{fig:7.7n} shows the variations of $E$ with $D_{dry}$, i.e. at first order, with the gel hydration level.


\textit{Fibers w/o droplets}, meaning fibers without inclusions, show smaller diameters than the one with inclusions. This is expected since the materials carried by the droplets is now missing. This effect is already visible while looking at the corresponding liquid structures, namely \textit{drops-in-jet} and \textit{jet$_1$}, as it can be seen on the central pictures of Fig. \ref{fig:7.2} or on those by Planchette et al. \cite{ref:Planchette2018}. Yet, the effect is much more pronounced once the fibers are left to equilibrate. Indeed, looking at Fig. \ref{fig:7.2} or considering continuity, the liquid section (or liquid diameter) ratio without and with droplets, is found to be in the range of 0.94 (or 0.97), much closer to 1 than the same ratio obtained comparing the section (or diameter) of the equilibrated fibers, namely 0.51 (or 0.71). This is probably caused by the composition of the droplet liquid, which contains, beside calcium cations, an important amount of glycerol (50\% in weight). Indeed, glycerol is known to be hygroscopic and its presence in the gel leads to an increased retention of water, similarly to what happens when more $Ca^{2+}$ adsorbed. 
Consequently and as expected, the value of $E$ is increased in the absence of droplets. With about 900$MPa$, it corresponds to the stiffest tested fiber. A more quantitative interpretation of this value is given at the end of this section while commenting the variations of $E$ with $D_{dry}$, see Fig. \ref{fig:7.7n}.

 
 \textit{Fibers w/o PEG}, i.e. fibers without PEG but with regular inclusions, show a significantly smaller diameter than the reference type with PEG. As already observed for the fibers left in a $Ca^{2+}$ bath and for those made without droplets, this decrease of diameter is accompanied by an increase of the Young's modulus. This might be attributed to the hygroscopic properties of PEG, which is at the origin of greater hydration level in equilibrated fibers \cite{ref:Baird2010}. Thus, in absence of PEG, the water retention decreases and, consequently, the Young's modulus increases. These results are also in agreement with several studies showing that using polyethylene glycol - or any hygroscopic plasticizer - leads to a reduction of  the Young's modulus \cite{ref:Wang2017, ref:Wongchitphimon2011, ref:Olivas2008, ref:Gao2017}. To better understand the effects at stake, it may be useful to look at the evolution of $E$ with $D_{dry}$. 
 
 Before discussing this point, let us briefly comment the results obtained by replacing $Ca^{2+}$ with $Sr^{2+}$. For this last fiber, we observe a smaller, yet comparable diameter of 100 $\mu m$ instead of 108 $\mu m$ for the reference fiber. The variation of the Young modulus is significant, increasing from 45 $MPa$ to 80 $MPa$ (more than 70\%) but remains in the same order of magnitude. Using fracture curves, Zhang et al. report more robust mechanical performance of strontium-alginate fibers compared to calcium-alginate ones \cite{ref:Zhang2020, ref:Morch2006}. Yet, the published stress-strain curves being not linear, no value of Young modulus is given preventing any direct comparison.
 In our case, we cannot exclude that the increase of $E$ obtained by using $Sr^{2+}$ is caused by the decrease of the fiber hydration, indirectly measured by $D_{dry}$. Indeed, the concurrent but opposed evolution of $E$ with $D_{dry}$ does not allow for immediate conclusion.

 To clarify this point, the previous data are plotted in Fig. \ref{fig:7.7n} in the form of $E$ versus $D_{dry}$. Interestingly, a string correlation between $E$ and $D_{dry}$ can be found for the fibers obtained with droplets and with PEG. The function $E=8~10^{27} {D_{dry}}^{-13}$, represented by the continuous grey line, satisfyingly represents this correlation. Thus, for these fibers, it seems that the variations of the Young modulus is primarily caused that the changes in $D_{dry}$, and thus, at first order, in the variations of the hydration level. This interpretation is in agreement with the results of Olivas et al. who showed that whatever the gel composition, their hydration level - tuned by conditioning under variable relative humidity - is the parameter influencing the most their mechanical properties. More particularly, the authors reported that the higher the hydration, the lower the Young modulus \cite{ref:Olivas2008}. Replacing $Ca^{2+}$ with $Sr^{2+}$ may lead to further variation of $E$ but it seems negligible compared to the effects of $D_{dry}$. 
 
The results obtained without PEG indicate that the Young's modulus does not increase as much as expected if considering only the evolution of $D_{dry}$. In other words, extrapolating the correlation found between $E$ and $D_{dry}$ to diameters smaller than 100 $\mu m$ (grey dashed line), slightly overestimates $E$. We do not have a clear explanation for this deviation but it remains rather small given the uncertainty of our measurements. Furthermore, other studies reporting that PEG increases the water retention and reduces the Young's modulus, do not aim in quantifying these effects, which limits possible comparisons \cite{ref:Wang2017, ref:Wongchitphimon2011, ref:Olivas2008}. Instead, this slight deviation brings us to question the validity of estimating  the fiber hydration level by $D_{dry}$.  Since less material is used in the absence of PEG,  $D_{dry}$ should be normalized by $D_{wet}$, the diameter of the fiber immediately after production. In our study, the reference fiber, those left in a liquid bath, and the one obtained replacing $Ca^{2+}$ with $Sr^{2+}$ are expected to have comparable $D_{wet}$. Yet,  the fibers obtained without PEG (or without droplets) are expected to have a reduced $D_{wet}$. Consequently,  estimating the hydration level with $D_{dry}/D_{wet}$ instead of $D_{dry}$ would mostly increase the hydration values obtained without PEG (or without droplets) and therefore reduce the mentioned deviation. 

In the case glycerol is missing, i.e. in the absence of droplets, a similar trend is observed: the measured value of $E$ is lower than its predicted value while extrapolating $E=8~10^{27} {D_{dry}}^{-13}$. Our interpretation is similar as for PEG. Glycerol is known to both increase the water retention of  alginate gel and to reduce its Young's modulus \cite{ref:Wongchitphimon2011, ref:Gao2017}. While Gao et al. \cite{ref:Gao2017} observe greater effects for PEG as for glycerol, we find an opposite trend. Yet, when present in our experiments, glycerol has a weight percentage 2.3 times larger than PEG, making this comparison irrelevant. Additionally and more importantly, the structure of the fiber with and without droplets is different. On one side, a fiber without inclusions is more regular, which is expected to lead to a more homogeneous stress distribution, and possibly to stiffer fibers. On the other hand, the droplets contain cross-linking ions and this additional gelation points main reinforce the overall mechanical properties of the fibers. To better evaluate the relative importance of these effects, new experiments could be performed using droplets containing PEG at adjusted concentrations to reach similar compositions with and without inclusions. This, however, goes beyond the purpose of the present study. Finally, it is important to keep in mind that our data has some uncertainty, which calls for careful interpretation rather than definitive conclusions.

 \subsection{Elongation at break}

\begin{figure}[t]
\centering
    \includegraphics[width=14cm]{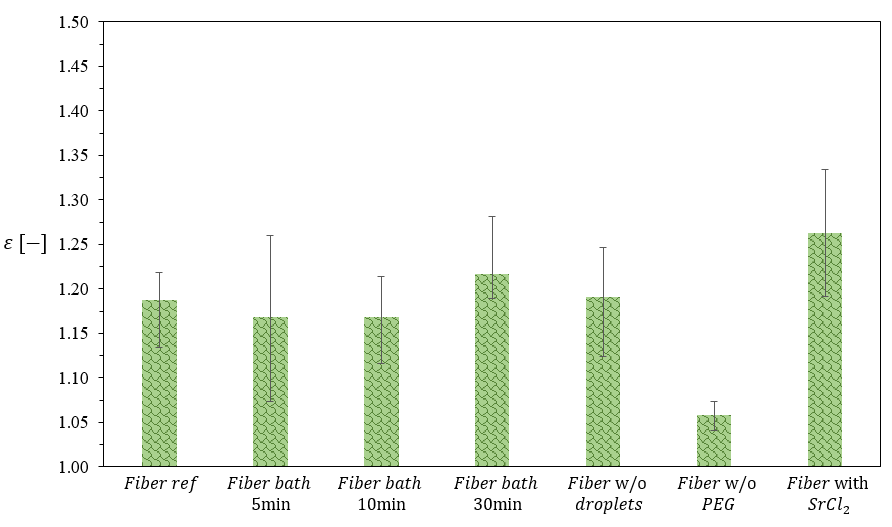}
  \caption{The elongation at break $\epsilon$ of the seven different types of fibers. For each type, the average value out of five samples is given. The error bars indicate the min and max values.}
  \label{fig:7.8}
\end{figure}

Let us now take a closer look to the elongation at break $\epsilon$ of the tested fibers. Here again, we test at least five samples of each type but in contrast to the Young modulus data, we keep the average value and not its maximum, see Fig. \ref{fig:7.8}. The error bars indicate the minimum and maximum values of each type of sample. As already mentioned, for these measurements, we only evaluate fibers which break in between the two clamping points, which automatically eliminates the possible errors mentioned for the Young Modulus and attributed to undetected damages at the clamping positions. Avoiding strong clamping, however, also induces the possibility that the fibers is held too weakly and thus slightly slide, producing another type of measurement error. Thus, we decide to apply the average value for the representation of $\epsilon$. 

A first look to Fig. \ref{fig:7.8} shows that the deviations between the different fiber types and the reference type are not as pronounced as for the Young's modulus. The breaking elongation of the reference fiber remains in the range of $18 \pm 5\%$, which is comparable to the breaking elongation of alginate-based fibers including PEG found in the literature. Wang et al. \cite{ref:Wang2007} obtained a breaking elongation between $19\%-24\%$ for sodium alginate fibers ($M_v=120000\,g/mol$) containing PEG6000 with mass fractions between $2\%-10\%$. Keeping the fiber in a liquid bath containing calcium chloride does not significantly increase this value. In contrast, replacing $Ca^{2+}$ by $Sr^{2+}$ increased the elongation at break up to $26 \pm 6\%$. The observed average breaking elongation therefore increases by a factor of 1.4 compared to the reference case. Similar results were obtained by Zhang et al. \cite{ref:Zhang2020} who reported an increase of the breaking elongation of pure sodium alginate fibers cross-linked with strontium ions by the factor of about 1.57 compared to fibers cross-linked with calcium ions. The effect is attributed to the stronger ionic bonding of the strontium cations with the carboxylic group of the alginate molecules, compared to calcium ions. It is therefore expected that not exactly the same factor is found with a different type of alginate \cite{ref:Zhang2020, ref:Morch2006}. Indeed, variations are known to be significant already while using the same type but changing from batch to batch. 

 The most dramatic effect on the breaking elongation is caused by the removal of PEG, see \textit{Fibers w/o PEG}. The elongation at break decreases down to values of about $5 \pm 2\%$. The absence of the plasticizer leads to more rigid fibers, which consequently break at smaller deformations. For comparable composition, similar values are reported in the literature where $\epsilon$ ranges between $4\%$ and $9\%$ \cite{ref:Wang2007, ref:Wang2019, ref:Zhang2020}. 

Finally, it seems that inclusions do not have any notable effect on the elongation at break as suggested by the values obtained for \textit{Fibers w/o droplets}. This surprising observation may result from two counter-playing effects that roughly compensate. As seen in Fig. \ref{fig:7.7} and reported in the literature \cite{ref:Olivas2008, ref:Gao2017}, glycerol has plasticizing effect and is therefore expected to increase the elongation at break of the fiber. Thus, its removal, which stems from the absence of droplets, should significantly reduce $\epsilon$. Simultaneously, this absence also leads to a more regular fiber, in which no inclusion are present. The latter can be seen as localized defects and certainly induces local stress inhomogeneities, which in turns triggers the rupture. Said differently, the absence of inclusions is expected to significantly increase the elongation at break. In practice, the absence of droplets does no noticeably affects the elongation at break. We thus make the hypothesis that the two previously mentioned effects more or less compensate each other. To go further, additional experiments are needed, which could for example used droplets but without any glycerol. This goes beyond the purpose of your work. 

\section{Conclusions}
\label{chap7:conclusions}
We have experimentally investigated, how fibers with and without regular inclusions can be produced via the in-flight collision of a regular droplet stream and two liquid jets. The throughput could be increased by an order of magnitude ($\approx 1500\,ml/h$) compared to the one obtained by other research groups working on fiber productions via in-air microfluidics. We designed a new experimental set-up, or in other words, a prototype, which allows, beside an online feedback loop on the droplets and jet alignment, to smoothly collect the generated fibers at velocities of $\approx 5\,m/s$. Note that, the available time span between the collision point of the three liquids and the collection plate is in the order of $200 \, ms$. Fairly small compared to the solidification time offered by other fiber production technologies, but enough to produce uniform fibers with comparable mechanical properties. In this study we tested seven different liquid combinations, whereby the diameters of the fibers let to equilibrate under the controlled laboratory conditions, were found between $73\, \mu m$ and $141\, \mu m$. To manufacture these fibers with regular inclusions, we used droplets with $D_d\approx 150 \mu m$ consisting of aqueous glycerol solutions containing $SrCl_2$ or $CaCl_2$, which collided with the first liquid jet. The jet was based on aqueous sodium alginate solutions, with and without polyethylene glycol, with a diameter of $\approx 320 \mu m $. The surface tension of the jet was slightly reduced using ethanol to enable the encapsulation of the droplets. The second jet with a diameter of $\approx 213 \mu m $ consisted of aqueous ethanol solutions also containing $SrCl_2$ or $CaCl_2$. To generate the fibers, the sol-gel transition of the sodium alginate jet was initiated from the inside, via the droplets, and from the outside, via the second jet, by means of ionic cross-linking triggered by divalent cations, here $Sr^{2+}$ and $Ca^{2+}$. After the collected fibers were left to equilibrate, their mechanical properties, namely Young's modulus $E$ and elongation at break $\epsilon$, were determined with the help of a measurement device manufactured in-house.     

The Young's modulus and the elongation at break of the reference fiber (\textit{Fiber ref}) were found to be $45 MPa$ and $18 \pm 5 \%$, respectively. This fiber was obtained using sodium alginate and PEG in combination with calcium chloride $CaCl_2$ as cross-linking agent in the droplets and the second jet. Placing similar fibers in a liquid bath containing the liquid of the second jet (aqueous ethanol solution + $CaCl_2$), however, drastically reduced the Young's modulus, about an order of magnitude, while the elongation at break did not seem to be significantly affected. This effect was attributed to the hygroscopic behaviour of calcium chloride, which led to highly hydrated fibers even after equilibration with the room conditions. Wet fibers generally show weaker mechanical properties than their dry counterparts. Sodium alginate fibers with PEG and without inclusions, as well as pure sodium alginate fibers with inclusions but without PEG, had the two greatest elasticity moduli obtained in this study, respectively close to $900 MPa$ and $460 MPa$. As expected, the absence of either glycerol or PEG, well known hygrophilic plasticizers, increased the Young's  modulus values. As in other studies, the concurrent decrease of the water retention, does not allow to draw any conclusion regarding possible  changes in intramolecular bonds involving  PEG or glycerol. The absence of either PEG or glycerol, however, did not have the same effects on the elongation at break. Without PEG but with inclusions, the elongation at break was found to be very small, around $5 \pm 2\%$, which can be explained by the absence of PEG's plasticizing effect. In case droplets were removed, no significant variation of the elongation at break was measured. This could be caused by two counter-acting effects, namely the lack of glycerol´s plasticizing action and the suppression of stress inhomogeneities probably developing around the inclusions. Finally, strontium chloride $SrCl_2$ was used for crosslinking the reference alginate solution containing PEG. In this case, the breaking elongation could be increased by a factor 1.40, while the increase of the Young modulus of a factor 1.77, rather seemed to be caused by the reduced hydration level.

\section*{Acknowledgements}

We would like to thank the Austrian Science Fund (FWF; grant no. P31064-N36) and the Austria Wirtschaftsservice Gesellschaft mbH (aws; patent no. EP 3412801 B1) for their financial support.

\end{document}